\documentclass[twocolumn,showpacs,floatfix,aps]{revtex4}
\usepackage{graphicx}
\usepackage{dcolumn}
\usepackage{bm}

\begin{document}

\title{Entanglement in quantum computers 
described by the XXZ model with defects}

\author{L. F. Santos}
\email{santos@pa.msu.edu} 
\affiliation{Department of Physics and Astronomy, Michigan State
  University, East Lansing, MI 48824}

\begin{abstract}
We investigate how to generate
maximally entangled states in systems 
characterized by the Hamiltonian of the XXZ model 
with defects. Some proposed quantum computers are described by 
such model. We show how the defects can be used to obtain
EPR states and $W$ states when one
or two excitations are considered.
\end{abstract}

\pacs{03.67.Lx, 03.65.Ud, 03.75.Gg, 75.10.Jm}

\maketitle

\section{Introduction}

Since qubits are two level systems, they are naturally 
modeled by spin-1/2 particles. 
Understanding spin chains is therefore very useful in the
study of quantum computers (QCs). 
Interaction between qubits corresponds then to 
interaction between spins.
One of the major problems of condensed matter based quantum computers
is that the interaction between qubits cannot be turned on and 
off when desired, and the quantum computer 
eigenstates soon become a linear 
superposition of a large number of
noninteracting multi-qubit states \cite{Shepelyansky}.
However, when performing computations, we would like to operate with well
defined states, in other words, we would like to entangle just
some specific states. In order to do so, we recur to an important
characteristic of most proposed QCs, that 
the energy difference between the qubits
states is large compared to the qubit-qubit interaction
and can be individually controlled \cite{2-5ourpaper,mark}.
In the QC based on electrons on helium, for example, the level 
spacing of each qubit is controlled by electrodes placed beneath
the helium surface \cite{mark}. 
The possibility of individually control qubit energies
allows us to entangle just some selected qubits by tuning 
them in resonance.

A fundamental requirement for the realization of quantum computation,
quantum teleportation and some protocols of quantum cryptography
is the generation of highly entangled quantum states. 
The maximally bipartite pure-state entanglement is identified
with the Bell or also called 
Einstein-Podolsky-Rosen (EPR) state $(1/\sqrt{2}) 
(|10\rangle + |01\rangle)$
\cite{popescu}. D\"ur {\it et al} showed that 
there are two different kinds of genuine tripartite
pure state entanglement: the 
maximally entangled Greenberger-Horne-Zeilinger (GHZ) state 
\cite{GHZ} and the so called $W$ state \cite{dur}. The 
$W$ state is the state of three qubits that retains
a maximal amount of bipartite entanglement when any one of the three 
qubits is traced out. It is written as 
$(1/\sqrt{3}) (|100\rangle + |010\rangle + |001\rangle)$.

There is a vast list of references dedicated to the problem of entanglement.
There have
been attempts to characterize qualitatively and quantitatively
the entanglement properties of multiparticle systems \cite{wootters}. 
Several proposals of how to prepare entangled states in different
kinds of systems have also been presented
\cite{quiroga}. Some papers 
investigated the entanglement between spins in a one
dimensional Heisenberg chain \cite{bose,wang}, which is
similar to what we intend to do here.
But contrary to these last cited papers, we 
consider a chain with defects and study their role in entangling states.
In a system where all qubits are in resonance but one, a defect
corresponds to the qubit whose level spacing is different from the others.

In this paper we investigate how to entangle selected
qubits in a system described by a strongly 
anisotropic one-dimensional XXZ model
with defects. This is the model used to describe the 
quantum computer based on electrons
on helium \cite{mark}. Since the coupling falls down quickly with 
the interqubit distance, we consider only nearest neighbor interaction.
The ground state of the system corresponds to all spins pointing down
and excitations correspond to spins pointing up. 
The interaction can only move excitations one site to the left or
to the right, so the number of excitations is constant.
We analyze systems with one or two spins up. In the 
case of one single excitation we show how the defects of the chain
can be used to maximally entangle two qubits. In the case of two 
excitations we show how EPR states and $W$ states can be obtained.

\section{Generation of maximally entangled states of selected qubits}

The Hamiltonian of the XXZ model with defects is

\begin{eqnarray}
\label{xxz}
&&H = \sum _{n=1} ^{N} \frac{\varepsilon _n}{2} \sigma _{n}^{z} + 
\frac{B}{2} \sum _{n=1} ^{N-1} \left[ \frac{\Delta }{2} 
\sigma _{n}^{z} \sigma _{n+1}^{z} + \frac{1}{2} H_{\mbox{hop}} 
\right] \\ 
&&H_{\mbox{hop}} = \left(\sigma _{n}^{+} 
\sigma _{n+1}^{-} + \sigma _{n}^{-} \sigma _{n+1}^{+}  \right) 
\nonumber,
\end{eqnarray}
where $\hbar =1$, $\sigma ^{z,+,-}$ are Pauli matrices, 
and $\varepsilon _n $ gives the energy difference between the two states
of qubit $n$. 
There are $N$ qubits. Here we 
consider a spin chain with free boundaries, which explains why 
the second sum runs over $n=1,..., N-1$.
In strongly anisotropic systems (such as the
QC based on electrons on helium), the parameter $\Delta $ 
is much larger than $B$. The last term in the Hamiltonian,
$H_{\mbox{hop}}$, is
responsible for the propagation of the excitation.

The ground state corresponds to all spins pointing down and its 
energy is ${\cal{E}}_0=-\sum_{n=1} ^{N}\varepsilon _n/2
+ (N-1)B\Delta /4$, which we will set equal to zero. Moreover, to
simplify our analysis we consider only positive values for the
parameters of the Hamitonian.

To address the different states of the system we use a notation that is
common in the study of spin chains with the Bethe ansatz \cite{us,bethe}.
The state corresponding to one single excitation on site $n$,
that is  $|\downarrow _{1} \downarrow _{2}...\downarrow _{n-1} 
\uparrow _{n} \downarrow _{n+1}... \downarrow _{N}\rangle $, is
simply written as $\phi (n)$. 
The state of two excitations, one on 
site $n$ and the other one on site $m$, is $\phi (n,m)$, 
which is a simplified notation for  
$|\downarrow _{1} ...\uparrow _{n} ...
\uparrow _{m} ... \downarrow _{N}\rangle $.

\subsection{One excitation}

Let us first examine the case of just one excitation.
Assume that there are only two defects, whose
level spacings are $\varepsilon _0 +g$, while the level
spacing of all other qubits is $\varepsilon _0$.
By choosing $g$ much larger than the interaction strength $B$, we
generate maximally entangled states corresponding to
linear combinations of the two defects.

The energy of states that have the excitation on 
any site but on the defects lies
within the band ${\cal{E}}_1\pm B$, where ${\cal{E}}_1 = 
\varepsilon _0 - B\Delta $ \cite{us,bethe}. If $g$ is much 
larger than $B$, an excitation on one of the defects will have
energy out of the band. Therefore the two resonant 
defects can be treated separately from all other
states by perturbation theory. 
It is as if we were working with only two sites.
An excitation initially created 
on one defect will only oscillate between the two defects. 
All intermediate states for the excitation to go from 
one defect to 
the other are virtual states. 
The frequency of these oscillations depend on the distance
between the two different qubits. Their separation determines
in which order of the perturbation theory they are connected.

The two states corresponding to superpositions of an
excitation on a defect on site $n_0$ and an excitation on
a defect on site $m_0$ are the following EPR states

\begin{equation}
\label{ent}
\psi _{\pm} = \frac{1}{\sqrt{2}} [\phi (n_0) \pm \phi (m_0)],
\end{equation}
where $\phi (n_0) = |\uparrow _{n_0} \downarrow _{m_0}\rangle $
and
$\phi (m_0) = |\downarrow _{n_0} \uparrow _{m_0}\rangle $

If the two defects are next to each other ($m_0=n_0+1$), 
the energies of these 
two entangled states in first order 
are 

\begin{equation}
E_{\pm} = {\cal{E}}_1 + g \pm B/2 
\end{equation}
We just need to diagonalize a 
two dimensional
sub matrix, whose diagonal elements are ${\cal{E}}_1 +g $ and 
off diagonal elements are $B/2$.
After shifting the energy levels by a second order correction
$B^2/4(g+B/2)$, they agree very well with a complete 
numerical diagonalization of a long chain with 
$g$ much larger than $B$. 

If the second defect is located on site $n_0+2$ we have to 
go straight to second order
and diagonalize a matrix whose diagonal elements are 
${\cal{E}}_1 +g +B^2/(2g)$ and off diagonal elements are
$B^2/(4g)$. The two energies become much closer,

\begin{eqnarray}
&&E_{+} = {\cal{E}}_1 + g + 3B^2/(4g) \\
&&E_{-} = {\cal{E}}_1 + g + B^2/(4g). \nonumber
\end{eqnarray}
The more distant the two defects are the closer the energies of the
two entangled states will be, since we have to go to higher 
orders to find them.

Suppose an initial state is prepared which has an excitation on the
defect $n_0$.
Let us now see how long we have to wait for it 
to become a maximally entangled state
such as given by Eq.(\ref{ent}). This 
excitation oscillates between the defect $n_0$ and the defect $m_0$.
The probability to find it later in time on site $n_0$ is

\begin{equation}
P_{\phi(n_0)} (t) = \frac{1 + \cos [(E_{+} - E_{-})t]}{2},
\label{T1}
\end{equation}
while the probability to find it on $m_0$ is

\begin{equation}
P_{\phi(m_0)} (t) = \frac{1 - \cos [(E_{+} - E_{-})t]}{2}.
\label{T2}
\end{equation}

It is seen from (\ref{T1}) and (\ref{T2}) that the period of oscillation
of the excitation between the defects is inversely proportional to the energy
difference of the states $\psi_{+}$ and $\psi_{-}$.
Such period depends on the number $\mu $ of sites between the defects
as $T_{\mu} = T_{0} (2g/B)^{\mu }$, where $T_{0} = 2\pi /B$.
The maximally entangled states (\ref{ent}) are obtained 
when $P_{\phi(n_0)}=P_{\phi(m_0)}=1/2$.
The further the defects are from each other the
longer we will have to wait for an EPR type of state to be created.
It is clear, however, that by tuning two qubits in resonance with 
energies very different from all the others, we can entangle
even remote qubits.

At the moment where a maximally entangled state is created, 
in order for it to be kept this way, the two defects have
to be detuned.
The difference in 
energy between these two excited qubits should become much 
larger than the interaction strength between them
(of course the defects and the other qubits are still completely 
out of resonance).
How fast this detuning should be done depends on how
much close we want to keep our state from a perfect EPR state.

Similarly, a $W$ state can be built with three resonant defects,
but we postpone the 
description of how to create such states to the next section where 
we have the more interesting case of two excitations.

\subsection{Two excitations}

In order to create EPR states and $W$ states with two excitations we
make use of the anisotropy of the system. Because $\Delta $ 
is much larger than $B$, two excitations next to each 
other have energy much larger than any state where they are
separated. The energy of any two free excitations lies inside 
the band $2{\cal{E}}_1 \pm 2B$, while the bound pair states have 
energy within a much narrower band 
$2{\cal{E}}_1 +B\Delta + B/2\Delta \pm B/2\Delta$ \cite{us,bethe}.
As a consequence, the bound pair states can be treated separately
from all other two-excitation states \cite{us}. They are connected in
second order of perturbation theory, which explains the 
narrow bandwidth
$B^2/(B \Delta )\equiv B/\Delta $. Any
intermediate and dissociated state is a virtual state.

Suppose that there is only one defect on site $n_0$ with level spacing 
$\varepsilon _0 +g$. If $g$ is much larger than $B/2\Delta $,
the bound pairs with one excitation on the defect have energy out of the
narrow band. They form the EPR states 

\begin{equation}
\psi _{\pm} = \frac{1}{\sqrt{2}} [\phi(n_0-1,n_0) \pm \phi (n_0,n_0+1)]
\end{equation}
where $\phi(n_0-1,n_0) = |\uparrow _{n_0 -1}\uparrow _{n_0 }
\downarrow _{n_0 +1} \rangle $
and 
$\phi (n_0,n_0+1) = |\downarrow _{n_0 -1}
\uparrow _{n_0 }\uparrow _{n_0 +1}\rangle $.
And their energies are

\begin{equation}
E_{\pm} = 2{\cal{E}}_1 +g +B\Delta + \frac{B}{4\Delta} +
\frac{B^2}{4(B\Delta +g)} \pm \frac{B^2}{4(B\Delta +g)}.
\end{equation}

By preparing an initial state with one excitation on site $n_0-1$ and
the other on the defect site $n_0$, 
following (\ref{T1}) and (\ref{T2}), we will obtain an EPR state
at every instant of time 

\begin{equation}
t_k = 2(B\Delta +g) [\pi/2 + k \pi]/B^2,
\end{equation}
where $k$ is an integer number.

Using the anisotropy, several other types of
EPR states can be created. With three defects on 
sites $n_0$, $n_0+1$ and $n_0+2$, if they all have the same level spacing
$\varepsilon _0 +g$ and $g\gg B/(2\Delta) $, we would
have linear combinations of the bound pairs 
$\phi (n_0,n_0+1)=|\uparrow _{n_0} \uparrow _{n_0+1} 
\downarrow _{n_0+2}\rangle $ 
and
$\phi (n_0+1,n_0+2)=|\downarrow _{n_0} \uparrow _{n_0+1} 
\uparrow _{n_0+2}\rangle $. 
As mentioned above, these states are
connected in second order of perturbation theory.

Another EPR state that can be 
created with those three defects involves the states $\phi (n_0,n_0+1)
=|\uparrow _{n_0} \uparrow _{n_0+1} \downarrow _{n_0+2}\rangle$  and
$\phi (n_0,n_0+2)=
|\uparrow _{n_0} \downarrow _{n_0+1} 
\uparrow _{n_0+2}\rangle $. The level spacing of the 
defect on site $n_0+2$ is now
$\varepsilon _0 +g +B\Delta$. The difference $B\Delta $ from the 
other two defects allows the entanglement between the
bound pair $\phi (n_0,n_0+1)$ and the state
$\phi (n_0,n_0+2)$. These states are connected in first
order in $B$.
The advantage of this entanglement is that
the period of oscillations between an initial state $\phi (n_0,n_0+1)$
and the state $\phi (n_0,n_0+2)$ 
is much shorter than oscillations between bound 
pairs. On the other hand, to 
guarantee that $\phi (n_0,n_0+1)$ and 
$\phi (n_0,n_0+2)$ are the only two states of the entanglement, $g$ has 
to be larger than $B$, instead of just larger than $B/(2\Delta)$ as in the
bound pair case.

This anisotropic chain with defects can also be used to 
create a $W$ state.
In order to do so we 
choose four equal defects on sites $n_0$, $n_0+1$, $n_0+2$ and $n_0+3$ with
level spacings $\varepsilon _0 +g$ and $g$ much larger than $B/(2\Delta )$.
The bound pairs on the defects are much higher in energy
than any other state and they can be treated separately.
It becomes a good approximation to say that three of the 
eigenfunctions of the total Hamiltonian (\ref{xxz}) will
correspond to linear combinations of the bound pairs
$\phi(n_0,n_0+1) = |\uparrow _{n_0} \uparrow _{n_0 +1}
\downarrow _{n_0+2} \downarrow _{n_0+3} \rangle$, 
$\phi(n_0+1,n_0+2)= |\downarrow _{n_0} \uparrow _{n_0 +1}
\uparrow _{n_0+2} \downarrow _{n_0+3} \rangle$ 
and 
$\phi (n_0+2,n_0+3)= |\downarrow _{n_0} \downarrow _{n_0 +1}
\uparrow _{n_0+2} \uparrow _{n_0+3} \rangle$. 
Such eigenfunctions
and their correspondent eigenvalues are obtained
from the diagonalization
of the following  tridiagonal sub matrix

\[
\left(
\begin{array}{ccc}
E^{(0)}+r+s & r &0 \\
r & E^{(0)} + 2r & r \\
0 & r & E^{(0)} + r + s
 \end{array}
\right) .
\]
Above $r=B/(4\Delta )$, $s = B^2/[4(B\Delta +g)]$
and $E^{(0)} = 2{\cal{E}}_1 + B\Delta + 2 g$.
The difference in energy between the diagonal element in the middle
of the matrix and
the diagonal elements at the edges, 
$B/(4 \Delta ) - B^2/[4\Delta (B\Delta +g)]$,
exists because state $\phi(n_0,n_0 +1)$ and $\phi(n_0+2,n_0 +3)$ make
virtual transitions to states that are out of the `sub' chain 
created by the defects.

The eigenvalues are therefore

\begin{eqnarray}
&&E_{1} = E^{(0)} +
\frac{4B^2 \Delta + 3Bg -B u}
{8 \Delta (B\Delta +g)} \nonumber \\
&&E_{2} = E^{(0)} +
\frac{B (2B\Delta +g)}{4\Delta (B\Delta +g)} \nonumber \\
&&E_{3} = E^{(0)} +
\frac{4B^2 \Delta + 3Bg +B u}
{8 \Delta (B\Delta +g)} \nonumber
\end{eqnarray}
where 
$u = \sqrt{8B^2 \Delta^2 + 16 B\Delta g + 9g^2}$. 

To generate a $W$ state we have to
prepare an initial state with excitations on $n_0+1$ and $n_0+2$, since this
state is the only one of the three which is 
connected to the two others in second order.
The probability in time to obtain the initial state $\phi(n_0+1,n_0+2) $
is 

\begin{equation}
P_{\phi(n_0+1,n_0+2)} (t) = \frac{1+\cos[(E_{3} - E_{1})t] }{2}.
\end{equation}
The probability to find state $\phi(n_0,n_0+1) $ 
or state $\phi(n_0+2,n_0+3) $ is the same and is given by

\begin{equation}
P_{\phi(n_0,n_0+1)} (t)  =
\frac{1-\cos[(E_{3} - E_{1})t] }{4},
\end{equation}
The $W$ state appears at 
the following instants of time (see Fig.1)

\begin{equation}
t_{k} = \frac{(-1)^k \arccos (-1/3) + 
2 \pi [k - \mbox{int}(k/2)]}{E_{3} - E_{1}}, 
\end{equation}
where $k$ is an integer and $\mbox{int}(k/2)$ 
correponds to the integer part of the ratio $k/2$.

\begin{figure}[h]
\includegraphics[width=3.2in]{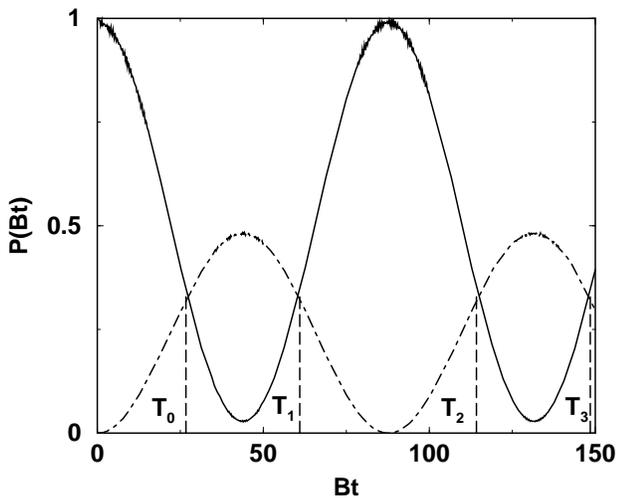}
\caption{We considered a chain with 12 qubits, $B=1$ and $\Delta =10$.
The defects are located on sites $n_0$, $n_0+1$, 
$n_0+2$ and $n_0+3$, where $n_0=3$ and $g=B\Delta $.
The solid line gives the probability in time to find the initial state
$\phi (n_0+1,n_0+2)$. Both probabilities 
to find state
$\phi (n_0,n_0+1)$ and state $\phi (n_0+2,n_0+3)$
coincide and they are given by the dot-dashed line. 
The vertical dashed lines
correspond to the instants of time, $T_{k} = B t_k$,
where we have a $W$ state.}
\end{figure}

As pointed out in the previous section, for 
the maximally entangled state to be kept this way, 
at the moment where it is generated, the level spacings
of the qubits involved in the process should become different.
This detuning should be larger than the interaction strength
among them.  

The $W$ state with the bound pairs $\phi(n_0,n_0+1)$, 
$\phi(n_0+1,n_0+2)$ and $\phi (n_0+2,n_0+3)$
can also be created with only two defects located 
on sites $n_0-1$ and $n_0+4$, but here the situation is more 
delicate. As before, $g$ has to be larger than the bandwidth of
the bound pair band, but it cannot be close to $B \Delta $,
because this would create resonances with states that have one 
excitation on the defect \cite{us}.

\section{Conclusion}

We have investigated how the defects of a spin chain 
with strongly anisotropic coupling can be used to
obtain EPR and $W$ states. These
are the states used in the study of 
bipartite and threepartite entanglement, respectively.
We considered the XXZ model with defects,
for this is the model used to
describe some quantum computers, such as the one
based on electrons on helium \cite{mark}.
It was shown that even though
the interaction among qubits
is on all the time, by controlling the level 
spacings we can determine among which qubits
the interaction is actually effective. This allows 
entangling only certain chosen sites.

Over the years, magneto-chemists have refined the art of designing 
and growing crystals of quasi-one-dimensional magnetic materials. 
These systems should therefore be useful in the study of
entanglement.

\acknowledgments 
We acknowledge support by the NSF through grant No. ITR-0085922 and
would also like to thank M. I. Dykman, 
C. O. Escobar and G. Rigolin for helpful discussions.

\end{document}